# Query Expansion Based on Clustered Results

Ziyang Liu     Sivaramakrishnan Natarajan     Yi Chen
Arizona State University
{ziyang.liu,snatara5,yi}@asu.edu

## ABSTRACT

Query expansion is a functionality of search engines that suggests a set of related queries for a user-issued keyword query. Typical corpus-driven keyword query expansion approaches return popular words in the results as expanded queries. Using these approaches, the expanded queries may correspond to a subset of possible query semantics, and thus miss relevant results. To handle ambiguous queries and exploratory queries, whose result relevance is difficult to judge, we propose a new framework for keyword query expansion: we start with clustering the results according to user specified granularity, and then generate expanded queries, such that one expanded query is generated for each cluster whose result set should ideally be the corresponding cluster. We formalize this problem and show its APX-hardness. Then we propose two efficient algorithms named iterative single-keyword refinement and partial elimination based convergence, respectively, which effectively generate a set of expanded queries from clustered results that provides a classification of the original query results. We believe our study of generating an optimal query based on the ground truth of the query results not only has applications in query expansion, but has significance for studying keyword search quality in general.

## 1. INTRODUCTION

While keyword search empowers ordinary users to search vast amount of data, delivering relevant results for keyword queries is challenging. Query expansion, or query refinement, is the process of reformulating a seed query to improve retrieval performance. Web search engines typically make query suggestion based on similar and popular queries in the query log [9, 2]. To handle a bootstrap situation where the query log is not available, there are works on query result summarization [24, 7, 5, 22, 15, 21], where popular words in the results are identified and suggested to the user for query refinement. The popularity of words are typically measured by factors such as term frequency, inverse document frequency, ranking of the results in which they appear, etc.

However, existing query expansion techniques based on result summarization using popular words can not effectively handle ambiguous queries which have multiple possible interpretations of their meanings, or exploratory queries [4] where the user does not have a specific search target, but would like to navigate the space of possibly relevant answers and iteratively find the most relevant ones. The expanded queries generated by such an approach may only cover a subset of the possible query semantics, and fail to provide a classification of the results. The problem becomes especially severe when the expanded queries are generated by summarizing the top-$K$ results, which is typically the case for efficiency reasons. One type of results may have higher ranks than other types and will suppress other result types to be reflected in the expanded queries. For instance, when searching "apple" on Google there is only one result about apple fruit in the top 30 results, whereas the rest are about Apple Inc. Since keywords about apple fruit have a small presence in these results, they are unlikely to be considered as popular words. Expanded queries generated according to popular words will bear the ranking bias and fail to cover the query semantics of searching apple fruit.

To handle ambiguous and exploratory queries, ideally query expansion should provide a *classification of different interpretations of the original query*, and thus guide the user to refine the query in order to get more results of the desirable type. For query "apple", intuitively, "Apple Inc." and "apple fruit" would be desirable, even though "fruit" is not a popular word in the set of top ranked results. Note that for this ambiguous query, either interpretation can be relevant to the user, although one interpretation is ranked much higher.

To generate expanded queries that provide a classification of the query results, we propose to first cluster the results into $k$ clusters[1] using one of the existing clustering methods, where $k$ is an upper bound specified by the user. In the above example of generating expanded queries for "apple" according to the top 30 results, although there is only one result about apple fruit, since it is significantly different from others, it should comprise a cluster itself, and thus can be covered by an expanded query.

Given a set of clusters of query results, the challenge is how to generate an expanded query for each cluster, whose set of results is as close to the cluster as possible. We assume that a result of a query is obtained by finding the data unit that contains all the query keywords. If we consider a cluster of results as the ground truth, our goal is then to generate a query whose set of results achieve both a high precision and a high recall. This is a difficult problem as the expanded queries should not only be selective to eliminate as many results in other clusters as possible (maximizing the precision), but also be general to retrieve as many results in this cluster as possible (maximizing the recall). One intuitive approach would apply existing works on cluster labeling / summarization [6, 19] to find the popular words in each cluster, and then use these words as the query for the cluster. However, the set of results retrieved by such a query would unlikely be similar to the original cluster. For example, consider 5 keywords, each appearing in 80% of the results in a cluster, but they do not co-exist in any result. A cluster labeling approach may output these 5 keywords as the label of the cluster. Nevertheless, using these 5 keywords as an expanded query will yield no result under AND semantics. This illustrates a unique challenge in generating queries for clustered results: the *interaction* of the keywords must be considered. Moreover, a potentially large number of results, and a large number of distinct keywords in the results add further challenges to the problem. Exhaustively searching for the optimal query for each cluster will be prohibitively expensive in practice. We formally define the problem of generating an optimal set of queries given the ground truth of each query's results. We show that this problem is NP-hard, and also APX-hard (i.e., it does not have a constant approximation).

---

[1] We can either cluster the set of all query results or a set of top ranked results.

To tackle the challenges, we propose two efficient algorithms. The first algorithm, named iterative single-keyword refinement (ISKR), iteratively refines a query in a greedy fashion by adding or removing a keyword to improve the quality of the query. The technical challenge is to dynamically and efficiently select the promising keywords to add to/remove from the current query. The second algorithm, named partial elimination based convergence (PEBC), attempts to find the best tradeoff between precision and recall using a randomized procedure. Specifically, given a set of sample queries and their F-measures, we find the two adjacent queries with the highest average F-measure, and iteratively test more points between them in search of an improved F-measure. Since the space of all possible queries is exponential to the data size, the technical challenge is how to efficiently find effective sample queries. We identify that this problem bears some similarity with the weighted partial set cover problem, but with fundamental differences that demand novel solutions. Compared to ISKR, PEBC in most cases favors more on the efficiency compared with quality, as shown in the experiments. Besides, when the results have ranking scores, both algorithms take the ranking scores into consideration by prioritizing the results with higher ranks when generating expanded queries.

The contributions of this work include:

- We propose a new philosophy for query expansion which aims at providing a classification of query results dynamically based on query results. This is especially useful for handling exploratory queries and ambiguous queries.
- We formally define the problem of generating optimal queries given the ground truth of the query results. This problem is APX-hard.
- To provide practical solutions to the query expansion problem, two algorithms which generate meaningful expanded queries efficiently given a clustering of the original query results have been proposed and developed.
- The quality and efficiency of the proposed approach are verified by experimental studies on real data sets.
- We believe the studies of generating an optimal query based on the ground truth query results not only has applications in query expansion, but has significance for studying keyword search quality in general.

The paper proceeds as follows. Section 2 introduces the problem definition. Two algorithms for generating expanded queries are presented in Sections 3 and 4, respectively, whose effectiveness and efficiency are empirically evaluated in Section 5. Section 7 concludes the paper with discussion of future works.

## 2. PROBLEM DEFINITION

In this paper, we consider keyword queries on either text documents or structured data. A text document is modeled as a set of words, and a structured document is modeled as a set of features defined as (entity:attribute:value) triplets [13], such as product:name:iPad.

Each result is a text document or a fragment of a structured document that contains *all* the keywords in the query.

The goal of this work is to generate a set of expanded queries that provides a classification of possible interpretations of the original user query. The input that we take includes a user query, and a set of clustered query results where the results are optionally ranked. Note that result clustering can be done using any existing clustering method (such as $k$-means), which is not the focus of this work. The output is one expanded query for each cluster of results, which maximally retrieves the results in the cluster, and minimally retrieves the results not in the cluster.

We now formally define the optimization goal. Considering the cluster as the ground truth, the quality of an expanded query can be measured using precision, recall and F-measure. Precision measures the correctness of the retrieved results, recall measures the completeness of the results, and F-measure is the harmonic mean of them. Let $C_1, \cdots, C_k$ denote the result clusters, $q_i$ denote the query generated for cluster $C_i$ ($1 \leq i \leq k$), $R(q_i)$ denote the set of results of $q_i$. The precision, recall and F-measure of $q_i$ are computed as

$$precision(q_i) = \frac{R(q_i) \cap C_i}{R(q_i)}, \quad recall(q_i) = \frac{R(q_i) \cap C_i}{C_i}$$

$$Fmeasure(q_i) = \frac{2 \times precision(q_i) \times recall(q_i)}{precision(q_i) + recall(q_i)}$$

To handle the general case where results are ranked, we use a weighted version of precision and recall. Let $S(\cdot)$ denote the total ranking score of a set of results, then

$$precision(q_i) = \frac{S(R(q_i) \cap C_i)}{S(R(q_i))}, \quad recall(q_i) = \frac{S(R(q_i) \cap C_i)}{S(C_i)}$$

The optimization goal is measured by the overall quality of the set of expanded queries (one for each cluster). We use the harmonic mean of their F-measures, whereas other aggregation functions (e.g., algebraic mean) can also be used.

$$score(q_1, \cdots, q_k) = \frac{n}{\frac{1}{Fmeasure(q_1)} + \cdots + \frac{1}{Fmeasure(q_k)}} \quad (1)$$

To summarize, the problem of generating expanded queries based on clustered results is defined as follows.

**Definition 2.1:** Given a set of clusters of query results, $C_1, \cdots, C_k$, retrieved by a user query under AND semantics, the *Query Expansion with Clusters* problem (QEC) is to find a set of queries, one for each cluster, such that their score (Eq. 1) is maximized. ∎

In fact, handling OR semantics is essentially the identical problem. More detailed discussion of OR semantics can be found in the Appendix.

The QEC problem is APX-hard. The proof is presented in our technical report [17].

In the next two sections we discuss the algorithms for query generation. Note that maximizing the overall score (Eq. 1) is equivalent as maximizing the F-measure of each query, thus each query can be generated independently. Specifically, the algorithms solve the following problem.

**Definition 2.2:** Given a user query $Q$, a cluster $C$ of results, and the set of results $U$ in all other clusters, as well as an optional ranking score of each result, the problem is to generate a query $q$, whose F-measure with $C$ as the ground truth is maximized. ∎

## 3. ITERATIVE SINGLE-KEYWORD REFINEMENT

The first algorithm we introduce is named *Iterative Single-Keyword Refinement* (ISKR). Given the user query and a cluster of results, the ISKR algorithm iteratively refines the input query until it cannot further refine the query to improve the F-measure of the query result (considering the cluster as the ground truth). Then, it outputs the refined query as the expanded query for the cluster. Specifically, it quantifies a value of each keyword appearing in the results, and refine the query by choosing the keyword with the highest value in each iteration. Several challenges need to be resolved for this approach to work: (1) How should we quantify and compute the value of a keyword? (2) As discussed in Section 1, keywords interact with each other when adding them to be part of a query. After the candidate query is refined, the value of a keyword may be changed. How should we identify the keywords whose values are affected and update the values of these keywords? (3) We start with the original user query, and try to add new keywords in the order of their values to this query to form an expanded query. Are there any case that a previously added keyword should be removed in order to improve the F-measure of the expanded query? (4) Since there can be a potentially large number of results and a large number of distinct keywords in a result, it is time-consuming to find the best set of keywords to add to the original query. How can we ensure efficiency? Next we will present ISKR algorithm, whose pseudo code can be found in the Appendix, that addresses these challenges.

**Value of a Keyword.** We first need to define the value based on which we choose the best keyword to add to or remove from $q$ at each step. When adding a keyword to a query $q$, the F-measure achieved by $q$ may either increase or decrease. Thus naturally, the value of a keyword should be measured by the delta F-measure of query $q$ after adding this keyword. But a disadvantage of this value function is that the values of the keywords are hard to maintain. The set of query results $R(q)$ is dynamically determined, based on the keywords that are already added to $q$. Since precision, recall, and thus F-measure are defined based on $R(q)$, the value of every keyword needs to be dynamically computed, and updated after every change to $q$.

To efficiently measure the values of keywords, we have the following observations. First, when adding a keyword $k$ to $q$, the positive effect is that $q$ may retrieve less results in $U$ (thus improving precision), and the negative effect is that $q$ may retrieve less results in $C$ (thus decreasing recall). Thus the number of results eliminated from $U$ and $C$ can be used to indicate whether it is good to add keyword $k$ to query $q$. Second, it is more efficient to maintain the number of results eliminated from $U$ and $C$ by adding a keyword $k$ than to maintain the delta F-measure of a keyword.

To see this, in the following, we use *delta results* of a keyword $k$ with respect to query $q$ (or simply delta results, if $k$ and $q$ are obvious) to denote the set of results retrieved by $q$, but not retrieved after adding $k$ to $q$. After adding $k$ to $q$, let $D$ denote the set of delta results. Consider a keyword $k'$ which appears in all results in $D$, i.e., it cannot eliminate any result in $D$. Note that the delta results of $k'$ with respect to query $q$ depends on how many results of $q$ can be eliminated by adding $k'$ to $q$. Since $k'$ cannot eliminate any result in $D$ anyway, the delta results of $k'$ with respect to $q$ are the same as the delta results of $k'$ with respect to $q \cup \{k\}$. This means that the delta results of $k'$ are not affected after adding $k$ to $q$.

With these observations, we measure the value of a keyword by benefit and cost. $benefit(k, q)$ is the total ranking score of the results eliminated in $U$ by adding $k$ to $q$, and $cost(k, q)$ is the total score of the results eliminated in $C$ by adding $k$ to $q$. Thus

$$benefit(k, q) = S(R(q) \cap U \cap E(k))$$

$$cost(k, q) = S(R(q) \cap C \cap E(k))$$

where $E(k)$ is the set of results that do not have keyword $k$ (hence will not be retrieved by any query that contains $k$).

We define the value of a keyword with respect to $q$ as its benefit-cost ratio, as commonly adopted in cost-benefit analysis:

$$value(k, q) = \frac{benefit(k, q)}{cost(k, q)} \quad (2)$$

We consider $value(k, q)$ as zero if both $benefit(k, q)$ and $cost(k, q)$ are zero.

**Identifying Keywords with Affected Values.** When we add a keyword to query $q$, the benefits and costs of other keywords may be affected. As discussed before, the value of a keyword is affected if and only if this keyword does not appear in at least one of the delta results. For each such keyword, we re-compute its benefit, cost and value using Eq. 2.

**Example 3.1:** We use this example to illustrate the ISKR algorithm. Suppose the original query is "apple". Consider a cluster $C$ with 8 results, $R_1, \cdots, R_8$, and $U$, the set of results that is not in $C$, with 10 results, $R'_1, \cdots, R'_{10}$. We consider 4 keywords for query expansion. The following table shows the keywords, and the results in $C$ and $U$ that each keyword can *eliminate*.

| | $k_i$ | $E(k_i) \cap C$ | $E(k_i) \cap U$ |
|---|---|---|---|
| $i = 1$ | job | $R_1, \cdots, R_6$ | $R'_1, \cdots, R'_8$ |
| $i = 2$ | store | $R_1, \cdots, R_4$ | $R'_1, \cdots, R'_4, R'_9$ |
| $i = 3$ | location | $R_2, \cdots, R_5$ | $R'_5, \cdots, R'_8, R'_{10}$ |
| $i = 4$ | fruit | $R_1, \cdots, R_3$ | $R'_2, \cdots, R'_4$ |

The initial benefit, cost and value of each keyword are:

| keyword | benefit | cost | value |
|---|---|---|---|
| job | 8 | 6 | 1.33 |
| store | 5 | 4 | 1.25 |
| location | 5 | 4 | 1.25 |
| fruit | 3 | 3 | 1.00 |

Since keyword *job* has the largest value, we first add *job* into $q$; so $q = \{$apple, job$\}$. Now $q$ retrieves 2 results in $C$: $R_7$ and $R_8$, and 2 results in $U$: $R'_9$ and $R'_{10}$.

Now we need to update the benefit, cost and value of each affected keyword. For example, the benefit of *store* becomes 1, since adding it to $q$ can further eliminate one result in $U$: $R'_9$. The cost of *store* becomes 0, since it does not eliminate any result in $C$, as both results ($R_7$ and $R_8$) contain *store*. The updated benefit, cost, and value of each keyword is shown in the following table (the row for *job* shows the benefit, cost and value of removing *job* from the current query, which will be discussed later).

| keyword | benefit | cost | value |
|---|---|---|---|
| job | 6 | 8 | 0.75 |
| store | 1 | 0 | ∞ |
| location | 1 | 0 | ∞ |
| fruit | 0 | 0 | 0 |

Thus we add *store* to $q$. After updating the benefit, cost, and value of the affected keywords, we further add keyword *location* to $q$. At this time, the only remaining keyword, *fruit*, has a value of 0, thus we do not further add keywords to the expanded query. ∎

**Necessity of Keyword Removal.** Since keywords added to the query may have complex interactions, it may be beneficial to remove a keyword from $q$ that was added to $q$ earlier, as shown in the following example.

**Example 3.2:** Continuing Example 3.1. Note that keyword *job* was added into $q$ at the first step due to its highest value, but after adding *store* and *location* to $q$, it becomes beneficial to remove *job*, which increases the recall but does not affect the precision. Indeed, the current $q = \{apple, job, store, location\}$ retrieves 2 results in $C$: $R_7, R_8$, and 0 result in $U$. If we now remove *job* from $q$, then $q$ will retrieve 1 more result in $C$: $R_6$, but still retrieve 0 result in $U$. Therefore, we should remove *job* from $q$ at this point. ∎

When removing a keyword $k \in q$ from $q$, the benefit, cost and value can be computed in a similar way. In contrast to keyword addition, removing $k$ from $q$ increases the results retrieved by $q$ in both $C$ and $U$, thus it may decrease the precision (measured by cost) and increase the recall (measured by benefit). For the removal case, the benefit and cost of $k$ with respect to $q$ are computed as

$$benefit(k,q) = S(C \cap D(k)), \quad cost(k,q) = S(U \cap D(k))$$

where $D(k)$ is the delta results after the removal of keyword $k$. $value(k, q)$ is still the benefit-cost ratio (Eq. 2).

Similar as adding a keyword, after removing a keyword, the values of other keywords may be affected. It is easy to see that the affected keywords are also those that do not appear in at least one of the delta results. For these keywords, we recompute their benefits, costs and values.

The ISKR algorithm stops when the query cannot be further improved by adding or removing a keyword, which is the case if the value of the best keyword is less than 1. In the running example, after updating the table, we find that no keyword has a value greater than 1, thus we stop and output the current query, $q = \{apple, store, location\}$.

## 4. PARTIAL ELIMINATION BASED CONVERGENCE

The ISKR algorithm iteratively attempts to add/remove a keyword to/from $q$, during which process the values of many keywords may change and need to be updated, which incurs a potentially high processing cost. In this section we propose a convergence based algorithm for query expansion named *Partial Elimination Based Convergence* (PEBC). It approaches the optimal solution in a fast and adjustable progress. Considering F-measure as a function over $q$, our goal is to find the value of $q$ that achieves the maximal value of F-measure. However, since the functional relationship between F-measure and $q$ is unknown, and the space of all possible queries is exponential to the data size, finding the optimal value is very challenging.

We propose algorithms that select several sample queries in the search space, and iteratively test more queries between the promising sample queries toward an improved F-measure. Specifically, given a set of queries and their F-measures, we find the two adjacent ones with the highest average F-measure, and test more points between them in search of an improved F-measure. The iteration continues until the expanded query is good enough, or enough iterations have been performed. The idea of this method is related to interpolation in numerical analysis, however, we do not infer the actual F-measure function from the sampled data points due to its high complexity.

Two questions must be resolved. (1) What type of sample queries we should use to converge to the optimal solution? (2) How can we obtain such sample queries?

**Type of sample queries.** To answer the first question, we propose to use a set of sample queries, each of which maximizes the number of results to be retrieved in $C$, given a percentage of results in $U$ to be eliminated. This is in the spirit of maximizing the recall given a fixed precision.[2] If we don't have the ranking scores of the results, we aim at eliminating $x\%$ of $U$'s results; otherwise, we aim at eliminating a set of $U$'s results, such that their total ranking score is $x\%$ of the total ranking score of all the results in $U$. In the following, we use "$x\%$ of the results in $U$" to refer to both cases.

**Example 4.1:** Suppose we generate five queries, $q_1$ to $q_5$, to eliminate 0%, 25%, 50%, 75% and 100% of the results in $U$, respectively, and maximize the number of results in $C$ to be retrieved. We compute the F-measures of these queries, and suppose they are: 0.5, 0.6, 0.4, 0.8, 0.1, respectively. Note that the F-measures of these queries may not have an obvious relationship. We take the two adjacent queries whose average F-measure is the highest, which are $q_3$ and $q_4$. We zoom in the interval between them, further dividing them to several intervals, and repeat the process. ∎

**Generating Sample Queries.** The key challenge of the PEBC algorithm is: given a percentage $x$ of results in $U$ to be eliminated, how can we generate query $q$ that eliminates roughly $x\%$ of the results in $U$, and maximizes the number of retrieved results in $C$? We refer to this problem as *partial elimination*.

This problem bears some similarity with the weighted partial set cover problem, which aims at using a set of subsets with the lowest total weight to cover at least $x\%$ of the elements in the universal set. However, in contrast to the partial weighted set cover problem which requires to cover *at least $x\%$* of the elements, our goal is to eliminate *as close to $x\%$ of the elements as possible*. This ensures that we can test data points that have roughly uniform distances between each other to better gauge the F-measure function. In the next subsections, we discuss how to address this new challenge and generate queries to achieve partial elimination.

### 4.1 Keyword Selection Based on Benefit/Cost

One intuitive method is to apply the greedy algorithm commonly used in weighted set cover for keyword selection: each time, we select the keyword with the largest benefit/cost ratio, until we have approximately $x\%$ of the results in $U$ eliminated. Benefit and cost are defined in the same way as in ISKR: benefit is the total weight of the un-eliminated results in $U$ that a keyword can eliminate, and cost is the total weight of the un-eliminated results in $C$ that a keyword can eliminate.

However, this method has an inherent problem that makes it infeasible: since the benefit/cost ratios of the keywords do not change with varying $x$, the keywords are always selected in the same order. Specifically, let the list of keywords selected when $x = 100$ be $\mathcal{K} = k_1, \cdots, k_p$. Now we want to select keywords to generate a query for each point in a range of possible values of $x$. No matter which point it is, the set of keywords selected will be a prefix of $\mathcal{K}$. Such a "fixed-order" selection of keywords makes it very difficult to control the percentage of results being eliminated.

---
[2]Alternatively, we can choose sample queries that maximize the number of results to be eliminated in $U$ given a percentage of results in $C$ to be retrieved.

**Example 4.2:** Consider a total of 10 results in $U$, $R_1, \cdots, R_{10}$, and 4 keywords: $k_1$=*job*, $k_2$=*store*, $k_3$=*location*, $k_4$=*fruit*. Suppose the set of results eliminated in $U$ by each keyword (benefit) and the number of results eliminated in $C$ by each keywords (cost) are:

$benefit(k_1) = 4(\{R_1, R_2, R_3, R_4\})$, $cost(k_1) = 2$
$benefit(k_2) = 6(\{R_5, R_6, R_7, R_8, R_9, R_{10}\})$, $cost(k_2) = 6$
$benefit(k_3) = 3(\{R_3, R_4, R_8\})$, $cost(k_3) = 1$
$benefit(k_4) = 4(\{R_4, R_5, R_6, R_7\})$, $cost(k_4) = 4$

Also suppose that the set of results in $C$ that is eliminated by a keyword does not intersect with the set eliminated by another keyword.

In this approach, the keywords are always selected in the decreasing order of their benefit/cost ratio, that is: $k_3 \rightarrow k_1 \rightarrow k_2 \rightarrow k_4$ (recall that after a keyword is selected, the benefit/cost of other keywords may change, as discussed in Section 3). Having the order of keyword selection fixed, there is a slim chance to achieve the goal of $x\%$ elimination. For instance, in order to eliminate 7 results with the fixed order keyword selection, we will have to either use $\{k_3, k_1\}$ which eliminates 5 results, or $\{k_3, k_1, k_2\}$ eliminating all 10 results. This poses a lot of restriction. Note that in this example, if we do not select keywords in this order, we can choose $\{k_1, k_4\}$ which eliminates exactly 7 results. ∎

As we can see, always selecting keywords based on their benefit/cost ratio makes it hard to eliminate a given percentage of the results. Next we discuss the approaches that overcome this problem using a randomized procedure.

## 4.2 Keyword Selection Based on a Selected Subset of Results

Since selecting keywords in a fixed order is undesirable, we propose to introduce a randomized procedure. First, we randomly select a subset of $x\%$ of the results in $U$. Then, we select the keywords, aiming at eliminating these randomly selected results. In this way, since the set of results to be eliminated is randomly selected, we will not select the keywords in a fixed order. If the randomly selected set of results is "good", we may be able to eliminate exactly this set of results.

Given the randomly selected results, selecting a set of keywords that eliminate these results with minimal cost is NP-hard, as the weighted set cover problem is a special case of it. To see this, assume that each keyword eliminates part of the selected set of results in $U$, and their costs are independent (i.e., they eliminates distinct sets of elements in $C$). Then, each keyword is equivalent to a subset in the weighted set cover problem. To choose a set of keywords that covers the randomly selected results, we can use some greedy approaches, e.g., let $S$ be the randomly selected set of results, at each time we choose a keyword which covers the most number of results in $S$ with minimal cost. Other methods can also be used.

**Example 4.3:** Continuing Example 4.2, suppose that we want to eliminate 7 results and the subset selected randomly is $\{R_1, R_2, R_3, R_4, R_5, R_6, R_7\}$. Given this set of results, we first update the benefits and costs of the four keywords. Keyword $k_1$ is not affected, as all four results it eliminates are selected. For $k_2$, we need to decrease its benefit by 3 because $R_8$, $R_9$ and $R_{10}$ are not selected, and increase its cost by 3. For $k_3$, we decrease its benefit and increase its cost by 1. $k_4$ is not affected. In this case, we can select $\{k_1, k_4\}$ which exactly eliminates this set of results.

However, if the randomly chosen subset is $\{R_1, R_2, R_3, R_4, R_8, R_9, R_{10}\}$, then the best we can do is: either using $k_1$ eliminating 4 of them, or using $\{k_1, k_2\}$ eliminating all 10 results. ∎

As we can see, this approach has two problems. First of all, given a set of randomly selected results, selecting a set of keywords that eliminate exactly this set of results with minimal cost is an NP-hard problem. Second, as illustrated in the above example, the quality of the algorithm highly depends on the selected subset, thus the chance that it can get the optimal answer is still slim.

## 4.3 Keyword Selection Based on a Selected Result

Both approaches discussed before put high restrictions on keyword selection, and thus generally suffer a low quality. We propose another randomized procedure that has a much better chance to eliminate as close to $x\%$ of the results in $U$ as possible. Instead of randomly selecting a subset of results, we randomly select *one* result in $U$ that is not eliminated yet, and then select a keyword that (1) can eliminate the selected result, (2) has the highest benefit cost ratio over all such keywords. In case of a tie, we choose the keyword that eliminates fewer results to minimize the risk that we eliminate too many results. If the percentage of the eliminated results is smaller than $x\%$, we continue the procedure; otherwise we stop and determine whether to include the last selected keyword based on which percentage is closer to $x\%$. Compared with the approach presented in Section 4.2, this one has a better chance of approaching the desired percentage, $x\%$, because selecting one result correctly is much easier than selecting a set of results correctly, as shown in Example 4.4.

**Example 4.4:** Continuing the example, to eliminate all 7 results, we may get the correct solution if we first choose one of the following five results: $R_1$, $R_2$, $R_5$, $R_6$ or $R_7$. Suppose that we choose $R_5$, and choose $k_4$ to eliminate it. After $k_4$ is used, we have the set $\{R_4, R_5, R_6, R_7\}$ eliminated. Then we can get optimal solution if the next randomly selected result is either $R_1$ or $R_2$. To eliminate $R_1$ or $R_2$, we choose $k_1$, which additionally eliminates results $\{R_1, R_2, R_3\}$, totaling 7 results eliminated. As we can see, the approach has a much higher chance to achieve the optimal solution (i.e. removing $x\%$ of results) than the ones discussed before. ∎

The pseudo code of the PEBC algorithm is shown in Algorithm 2 (Appendix).

## 5. EXPERIMENTS

In this section we report a set of experimental evaluations on the quality of the expanded queries generated by our approach, and the efficiency and scalability of query generation.

### 5.1 Experimental Setup

We tested our approaches on two data sets: shopping and Wikipedia. We used 20 queries, 10 for each data. We compared our approach with four existing methods for query expansion: Data Clouds [15], Cluster Summarization (CS) [6], Google, and a variation of ISKR where the value of a keyword is considered as the change of F-measure after adding/removing the keyword. The detailed experimental setup and the description of comparison systems can be found in the Appendix.

### 5.2 Quality of Query Expansion

The evaluation of the quality of expanded queries consists of a user study and the measurement of scores of the expanded queries (Eq. 1).

#### 5.2.1 User Study

We performed a user study an Amazon Mechanical Turk and had 45 users participate in our survey for evaluating the query expansion approaches. The user study consists of three parts.

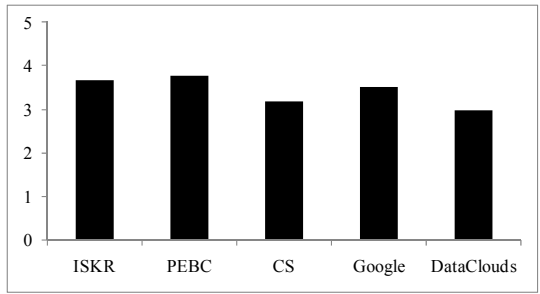

Figure 1: Average Individual Query Score

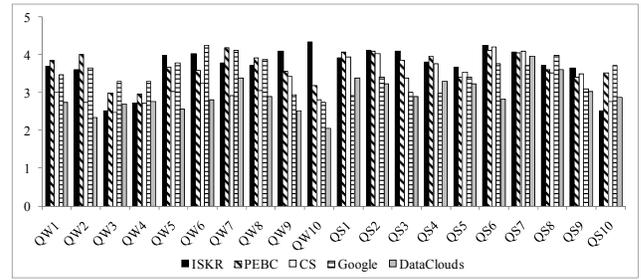

Figure 3: Collective Query Score for Each Set of Expanded Queries

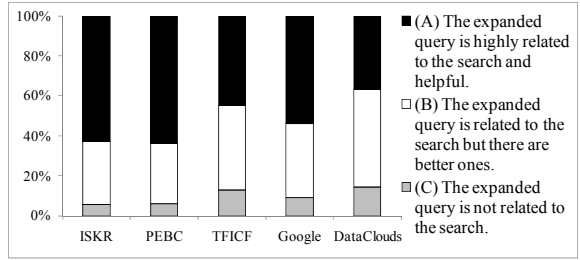

Figure 2: Percentage of Users Choosing Options (A), (B) and (C) for Individual Queries

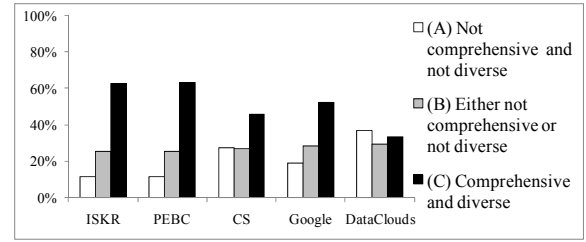

Figure 4: Percentage of Users Choosing Options (A), (B), (C) and (D) for Each Set of Expanded Queries

**Part 1: Individual Query Score.** In order to test whether an expanded query generated by each approach is helpful for the users, we first asked the users to give a score for each expanded query in the range of 1-5, which is referred to as *individual query score*. The users were also asked to choose one of the options shown in Figure 2 as the justification of the score.

The average score of all 20 queries given by all users for each approach is shown in Figure 1, and the percentage of users choosing each option in this part of the user study is shown in Figure 2. As we can see, ISKR, PEBC and Google have higher average query scores than Data Clouds and CS. Recall that data clouds returns the top keywords in all results in terms of tf, idf and result rank, but such a top keyword may often be too specific (e.g., "multicellular" for $QW_7$) or too general, which is not informative as an expanded query. On the other hand, an expanded query generated by ISKR and PEBC maximally retrieves a cluster of results, thus is likely to have a better semantics. For example, for $QW_6$ ISKR and PEBC return "island" and "server", which are more meaningful. Therefore, most users chose option (A) for both ISKR and PEBC, while data clouds got plenty of (B) and (C).

The CS approach chooses keywords based on TFICF, thus may tend to pick keywords that have high occurrence (TF) in a few results in the cluster. These keywords do not retrieve many results in the cluster, thus the users mostly found it less desirable. For example, for $QW_6$ "java", it returns a query "Java, blog, Microsoft", which is too specific and only cover a small part of the results.

Google chooses keywords based on query log, thus it often returns meaningful and popular keywords. For example, for $QW_6$ "Java", Google returns the expanded queries "Java, Tutorials", "Java, Games" etc., which are generally very popular with the users. However, for some queries Google may return keywords that do not occur in the results. For example, for $QS_1$ "Canon, products", Google returns a query "Sony, products". While this could be useful for some users, our user rating has indicated that many the users prefer the expanded queries to be results oriented.

There are a few queries that ISKR and/or PEBC do not generate the most meaningful expanded queries. This is mainly because the words that appear frequently in a cluster is not necessarily the best one semantically. It is especially likely for the Wikipedia data set, as it consists of document-centric XML with sentences/paragraphs, rather than succinct and informative features. Consider $QW_1$ "San Jose", for which one of our expanded queries is "player". Although this keyword is related to the sports teams of San Jose, the users suggested that returning team information (e.g., baseball, hockey, etc.) gives better expanded queries.

**Part 2: Collective Query Score.** Next we test whether the *set* of expanded queries for each user query provides a classification of the original query result set. We asked the users to give a collective score for all expanded queries of each user query, in the range of 1-5, and choose one of the options in Figure 4 as the justification of the score.

For all 20 queries, the collective score of each user query for each approach is shown in Figure 3, and the percentage of users that chose each option is shown in Figure 4. As we can see, ISKR, PEBC consistently received relatively high scores in collective scoring. Since each expanded query of ISKR and PEBC maximally covers the results in a cluster and minimally retrieves the results in other clusters, they are usually comprehensive (i.e., covering various aspects/meanings of the original query) and diverse (i.e., their results have little overlap). This was appreciated by the users as it is easy for the users to see all options and decide the expanded query for retrieving the relevant results, and the users gave favorable scores for ISKR and PEBC.

On the other hand, since Data Clouds only returns the top keywords in the results, the expanded queries may lack comprehensiveness and diversity. Consider $QS_1$ "Canon, products". Both ISKR and PEBC returns three main products of Canon: *camera*, *printer* and *camcorder*. However, data clouds returns *camera*, *printer* and *wp-dc26* (a camera-related product). As we can see, the expanded queries of data clouds do not cover camcorder products, failing to be comprehensive. Besides, the result of *wp-dc26* is contained in the result of *camera*, failing to be diverse. As a result, the users mainly chose options (A) and (B) for data clouds.

For the CS approach, as discussed before, it tends to pick keywords that have high occurrence in fewer results and do not cover the entire cluster. Such queries are usually too specific and thus fail to be comprehensive. For example, for $QW_1$ "San Jose", CS returns "San Jose, sabercat, season, arena" and "San Jose, war, California, gold". Since these expanded queries only retrieve a few results in the corresponding clusters, the user found them not comprehensive. Note that the CS approach has a better score on shopping data than the Wikipedia data. This is because in the shopping data, results are somewhat similar in that they share many common keywords. Therefore, even though the CS approach does not consider the relationship of keywords, the keywords it selects in an expanded query likely co-occur in many results. On the other hand, on the Wikipedia data, it may choose a set of keywords, such that each of them has a high occurrence but they do not necessarily co-occur. Such a query will not retrieve many results which lowers its recall. For example, for $QW_9$ "mouse", it returns a query "mouse, technique, wheel, interface". These keywords have high occurrences but low co-occurrences.

Since Google chooses expanded queries based on query log, it can also achieve comprehensiveness and diversity for some queries. For example, for $QW_6$ "Java", Google returns the expanded queries "Java, Tutorials", "Java, Games" and "Java, test", which the users considered as comprehensive and diverse. However, sometimes the expanded queries returned by Google may not be diverse. For example, for $QW_8$ "rockets", all expanded queries returned by Google are about space rockets, and none of them refers to the Rockets NBA team.

There are also a few cases where users choose (A) or (B) for ISKR and/or PEBC. Due to the limitation of the data, we may not have the results that cover all meanings of a query in the results. As an example, consider query $QW_1$ "San Jose", the top-30 results are either about the city of San Jose in California, or about San Jose sports teams. Since San Jose is also a major city in Costa Rica, which is not covered by our expanded queries, some users selected (A) or (B) for our approach. Besides, due to imperfect clustering, sometimes it may be impossible to generate comprehensive and diverse expanded queries.

**Part 3.** To verify the intuition of our approach, we finally asked the users a general question: What is your opinion about a good set of expanded queries? According to the answers from the users, the majority of users considered comprehensiveness and diversity as important properties for a set of expanded queries, which coincides with the philosophy of our proposed approaches.

*5.2.2 Scores of Expanded Queries Using Eq. 1*

As defined in Eq. 1, the score (goal function) of a set of expanded queries is the harmonic mean of their F-measures. In this section we test for each user query the score of expanded queries generated by ISKR, PEBC, the F-measure approach and CS, as shown in Figure 5. Since the queries generated by Data Clouds and Google are not based on clusters, this score is inapplicable.

As we can see, in general ISKR and PEBC achieve similar and good scores. On the shopping data, both algorithms achieve perfect score for many queries. This is because on the shopping data, products of different categories usually have different features. Thus for queries whose results contain several different product categories (e.g., $QS_1$ "Canon, products" whose products contain camcorders, printers, and cameras), each category forms a cluster, and it is usually possible to achieve a perfect precision and recall.

The scores of ISKR are generally a little better than those of PEBC. The reason is that in each iteration of ISKR, we select the best keywords to add to $q$ or remove from $q$. Thus ISKR, although not necessarily produces the optimal expanded queries, does achieve some form of local optimality: it stops only if no single keyword can give a better value if we add it to $q$ or remove it from $q$. On the other hand, PEBC relies on the assumption that, if two adjacent points have the highest average score, then the optimal query should lie in between these two points. Since this assumption is not always true, sometimes PEBC may not choose the best interval to zoom in. However, if PEBC chooses the right interval at each iteration, then it may achieve a better quality than ISKR as it will converge to the optimal solution, as the case of $QS_4$, $QW_{10}$, etc.

The F-measure approach generally has the same or slightly better quality than ISKR since delta F-measure is a more accurate measure of the value of a keyword. For some queries its scores are lower, since both algorithms are heuristics-based and ISKR may occasionally choose better keywords. However, as shown in Section 5.3, the F-measure approach has a poor efficiency, while efficiency is highly important for a search engine.

The CS approach usually has a poor score. This is because it chooses a set of keywords with high TFICF values with respect to a cluster, but these keywords may not occur in many results in the cluster, thus causing a low recall. For example, for $QW_5$ "eclipse" it returns a query "eclipse, core, plugin, official". Moreover, as discussed before, since the CS approach is designed to return cluster labels rather than query results, it does not consider the interaction of keywords. Therefore, it may return an expanded query whose keywords have high occurrences, but low co-occurrences.

### 5.3 Efficiency and Scalability

In the efficiency test, we measure the running time of all five methods. For all approaches except Data Clouds, the response time that the user perceives (besides the query processing time) includes both clustering and query expansion time. The average clustering time on shopping and Wiki data sets are 0.02s and 0.35s, respectively. For data clouds, we measure the time for finding the top-$k$ words from a ranked list of results.

The processing time of query expansion is shown in Figure 6. In general, the ISKR algorithm takes more time to generate expanded queries comparing with PEBC. Recall that ISKR allows both adding or removing keywords to the current expanded query, thus it may have a large number of iterations before getting to the point that the query cannot be further improved. Besides, at each iteration ISKR needs to maintain the values of keywords, and in the worst case, the values of all keywords need to be updated. For $QS_8$ which has a large number of results (557) and a large number of distinct keywords (464 in the largest cluster), it is significantly slower than PEBC.

Both ISKR and PEBC are much more efficient than the F-measure method. As discussed in Section 3, the F-measure method needs to update the values (i.e., delta F-measure) of all keywords every time a keyword is added to or removed from a query. On the other hand, ISKR only needs to update the values of the keywords that do not appear in all delta results, the number of which could be a small percentage of all distinct keywords. For some queries the F-measure method takes more than 30 seconds. The efficiency of the CS approach is usually comparable with ISKR and PEBC, since the TFICF of a keyword can be efficiently computed. Data clouds is generally faster than both ISKR and PEBC, as it only needs to compute the tf and idf for each keyword in the results.

The scalability test is presented in the The Appendix.

### 6. RELATED WORK

There are a number of different existing approaches on query expansion, e.g., there are works based on query log [2, 9], general or domain-specific ontology [3, 12, 10], user profile and collaboratively filtering [11], selecting popular words in the results [24, 7, 5, 22, 15, 21, 23], etc. Query expansion is also related to the topics of

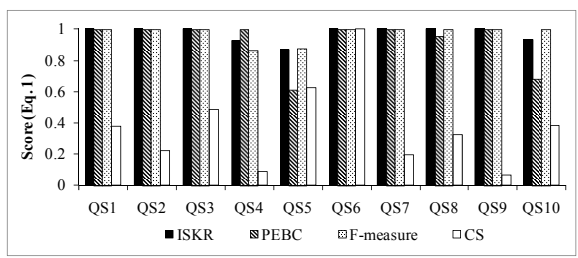
(a) Shopping Dataset

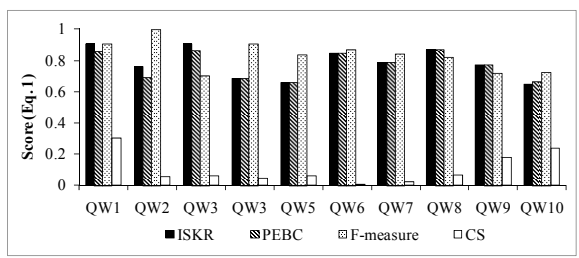
(b) Wikipedia Dataset

**Figure 5: Scores of Expanded Queries (Eq. 1)**

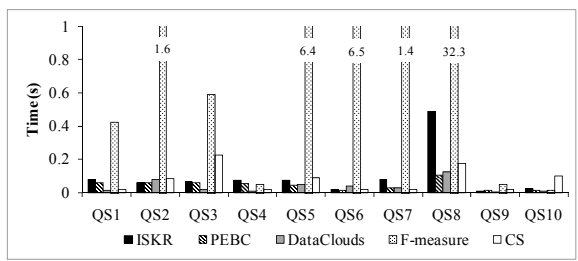
(a) Shopping Dataset

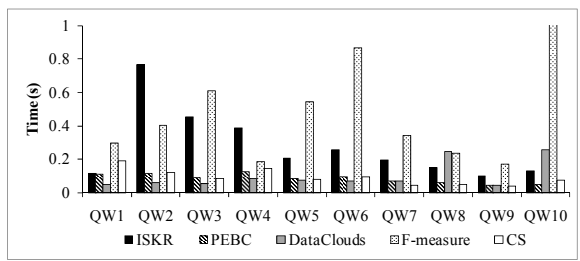
(b) Wikipedia Dataset

**Figure 6: Query Expansion Time**

faceted search [8, 14, 16], cluster labeling/summarization [6, 19], result differentiation [18], etc. More detailed discussion of related works is presented in the Appendix.

## 7. CONCLUSIONS AND FUTURE WORK

In this paper we propose a novel framework for query expansion: generating a set of expanded queries that provides a classification of the original query result set. Specifically, the expanded queries maximally retrieve the results of the original query, and the results retrieved by different expanded queries are different. To achieve this, we propose to first cluster the results, and then generate an expanded query for each cluster, whose set of results should be as close to the cluster as possible. We formally define the Query Expansion with Clusters (QEC) problem. This problem is APX-hard. We then design two efficient algorithms ISKR and PEBC for generating expanded queries based on the clustered results. In the future, we would like to investigate how different clustering methods affect the expanded queries, and design techniques for choosing the best clustering method dynamically. We would also like to study how to support vector space retrieval model, as well as the possibility of interweaving the clustering and query expansion process.

## 8. ACKNOWLEDGEMENT

This material is based on work partially supported by NSF CAREER Award IIS-0845647, IIS-0915438 and IBM Faculty Award.

# APPENDIX

## A. Pseudo Code of Algorithm 1

**Algorithm 1** Iterative Single-Keyword Refinement

ISKR (User Query: $uq$, Cluster: $C$, Results not in $C$: $U$)
1: $\mathcal{K}$ = the set of keywords in $C \cup U$
2: $q = uq$
3: $Refine(C, U, \mathcal{K}, q, weight)$
4: return $q$

REFINE $(C, U, \mathcal{K}, q, weight))$
1: $T = \emptyset$
2: **for** each $k \in \mathcal{K}, k \notin q$ **do**
3:    $E(k)$ = the set of results that do not contain $k$
4:    $benefit(k) = S(R(q) \cap C \cap E(k))$
5:    $cost(k) = S(R(q) \cap U \cap E(k))$
6:    $value(k) = benefit(k)/cost(k)$
7:    insert $k$ into $T$
8: **end for**
9: **for** each $k \in \mathcal{K}, k \in q$ **do**
10:    $D(k) = R(q\backslash k)\backslash R(q)$
11:    $benefit(k) = S(D(k) \cap C)$
12:    $cost(k) = S(D(k) \cap U)$
13: **end for**
14: **while** true **do**
15:    $k$ = top-1 keyword in $T$
16:    **if** $value(k) \leq 1$ **then**
17:      break
18:    **end if**
19:    **if** $k \in q$ **then**
20:      $q = q\backslash k$
21:      $MaintainT(T, q, k, E(k), \mathcal{K}, C, \text{remove})$
22:    **else**
23:      $q = q \cup k$
24:      $MaintainT(T, q, k, E(k), \mathcal{K}, C, \text{add})$
25:    **end if**
26: **end while**
27: return $q$

MAINTAINT $(T, q, k, E(k), \mathcal{K}, C, type)$
1: **if** $type$ = add **then**
2:    $deltaResult = R(q\backslash k) \cap E(k)$
3: **else**
4:    $deltaResult = R(q\backslash k)\backslash R(q)$
5: **end if**
6: **for** each $k' \in \mathcal{K}$ **do**
7:    **if** each $k'$ appears in all results in $deltaResult$ **then**
8:      continue
9:    **end if**
10:    **if** type = add **then**
11:      $benefit(k') = R(q) \cap U \cap E(k')$
12:      $cost(k') = R(q) \cap C \cap E(k')$
13:    **else**
14:      $D(k) = R(q\backslash k)\backslash R(q)$
15:      $benefit(k') = D(k) \cap C$
16:      $cost(k') = D(k) \cap U$
17:    **end if**
18:    remove $k'$ from $T$
19:    $value(k') = benefit(k')/cost(k')$
20:    add $k'$ to $T$
21: **end for**

## B. Pseudo Code of Algorithm 2

**Algorithm 2** Partial Elimination Based Convergence

PEBC (User Query: $uq$, Cluster: $C$, Results not in $C$: $U$)
1: $\mathcal{K}$ = the set of keywords in $C \cup U$
2: $q = uq$
3: $Converge(C, U, \mathcal{K}, q)$
4: return $q$

CONVERGE $(C, U, \mathcal{K}, q))$
1: $nseg$ = 5 {set the number of segments to split the interval}
2: $nit$ = 5 {set the number of iterations}
3: $left = 0, right = 100, step = (right - left)/nseg$
4: **for** $i$=1 to $nit$ **do**
5:    **for** $x = left; x \leq right; x+ = step$ **do**
6:      $currC = C, currU = U$
7:      **repeat**
8:         $r$ = a randomly selected result
9:         $bestvalue = 0$
10:        **for** each distinct keyword $k \notin r$ **do**
11:           $E(k)$ = the set of results that do not contain $k$
12:           $benefit(k) = E(k) \cap U$
13:           $cost(k) = E(k) \cap C$
14:           $value(k) = benefit(k)/cost(k)$
15:           **if** $value(k) > bestvalue$ **then**
16:              $selecetd = k, bestvalue = value(k)$
17:           **end if**
18:        **end for**
19:        $q = q \cup selected$
20:        $currC = C\backslash E(k), currU = U\backslash E(k)$
21:      **until** roughly $x\%$ percent of results in $U$ are eliminated
22:    **end for**
23:    $left, right$ = the interval with the largest average score
24: **end for**

## C. Experimental Setup

**Environment.** All experiments were performed on a machine with AMD Atholon 64 X2 Dual Core Processor 6000+ CPU with 3GHz, 4GB RAM, running Windows Server 2008.

**Data Set.** We tested our approaches on two data sets: shopping and Wikipedia. Shopping is a data set that contains information of electronic products crawled from circuitcity.com. Each product has a title, a category, and a set of features. Wikipedia is a collection of document-centric XML files used in INEX 2009.[3]

**Query Set and Result Clustering.** We tested 10 queries on each data set, as shown in Table 1 (Appendix). The queries on Wikipedia dataset are composed of ambiguous words. The queries on shopping dataset are to search for specific products. We adopt $k$-means for result clustering. Each result is modeled as a vector whose components are features in the results and the weight of each component is the TF of the feature. The similarity of two results is the cosine similarity of the vectors.

**Comparison Systems.** We compared the proposed ISKR and PEBC algorithms with several representative query expansion methods:

(1) *Data Clouds* [15], which takes a set of ranked results, and returns the top-$k$ important words in the results. The importance of a word is measured by its term frequency in the results it appears, inverse document frequency, as well as the ranking score of the results that contain the word. Data Clouds is a representative method for returning important words in the search results, without clustering the results.

(2) *Google*. For each test query, we take the first 3-5 related queries

---
[3] http://www.inex.otago.ac.nz/

suggested by Google (the number of which is the same as the number of queries generated by other approaches). Google is a representative work of suggesting related queries using query logs.

(3) *CS*, representing Cluster Summarization [6]. It first clusters the results, then generates a label for each cluster. The label of a cluster is selected based on the term frequency (tf) and inverse cluster frequency (icf) of the words in the cluster. CS is a representative method for cluster summarization and labeling.

(4) *F-measure*, which is an alternative ISKR algorithm that considers the value of a keyword $k$ with respect to a query $q$ as the delta F-measure of $q$ after adding $k$ to $q$ or removing $k$ from $q$. As discussed in Section 3, since our goal function is to maximize the F-measure of a query, the delta F-measure more accurately reflects the value of a keyword than the benefit/cost ratio. However, in this approach, after a keyword is added to or removed form the current query, the values of all keywords will need to be updated, which potentially leads to a low efficiency.

We implemented Data Clouds and CS.

In ISKR and PEBC, we consider the top-20% words in the results in terms of tfidf for query expansion. In PEBC, we empirically set the number of points tested in each iteration as 3, and the number of iterations as 3. Since there are a lot of results for queries on Wikipedia data set, all systems only consider the top 30 results to generate expanded queries, where the results are ranked using tfidf of the keywords. We also set the maximal number of expanded queries for each approach to be 5.

### D. Test Queries

The test queries are shown in Table 1.

| Wikipedia | |
|---|---|
| $QW_1$ | San Jose |
| $QW_2$ | Columbia |
| $QW_3$ | CVS |
| $QW_4$ | Domino |
| $QW_5$ | Eclipse |
| $QW_6$ | Java |
| $QW_7$ | Cell |
| $QW_8$ | Rockets |
| $QW_9$ | Mouse |
| $QW_{10}$ | sportsman, Williams |
| Shopping | |
| $QS_1$ | Canon Products |
| $QS_2$ | Networking Products |
| $QS_3$ | Networking Products Routers |
| $QS_4$ | TV |
| $QS_5$ | TV Plasma |
| $QS_6$ | HP Products |
| $QS_7$ | Memory |
| $QS_8$ | Memory 8GB |
| $QS_9$ | Memory Internal |
| $QS_{10}$ | Printer |

**Table 1: Data and Query Sets**

### E. Scalability of Query Expansion

We have tested the scalability of all three approaches with respect to the number of results returned by the user query. We use query $QW_2$ "Columbia", and vary the number of results from 100 to 500. The time shown for ISKR and PEBC include both clustering and query generation. As shown in Figure 7, the processing time of both approaches increases linearly, and they have a reasonable response time even if 500 results are used.

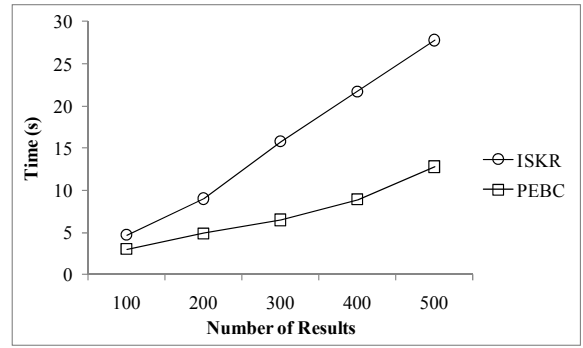

**Figure 7: Scalability over Number of Results**

### F. Related Work

**Query Expansion.** Expanded queries can be generated based on query log [2, 9], general or domain-specific ontology [3, 12, 10], user profile and collaboratively filtering [11]. Since such information may not always be available, there are also works that generate expanded queries only based on the information contained in the corpus, i.e., the results retrieved by the user query and/or the entire data repository. As our work falls into this category, we focus the discussion on corpus-driven query expansion.

One class of work is based on relevance feedback. The expanded query aims to retrieve a set of results that are similar to the relevant results, where the relevant results are specified by the user in explicit feedback or are considered to be the top ranked results in pseudo feedback. To generate new queries, various approaches have been proposed to select and rank terms from relevant results, including TF-IDF based methods [15, 24], probabilistic language model based methods [20], vector space model based methods [25], etc. However, since users typically provide feedback to top ranked results only, top ranked results are mostly likely reinforced and the diversity of the results are compromised. Furthermore, the pseudo feedback approach assumes that relevant documents are similar to each other, and are quite different from irrelevant ones. Thus relevance feedback approach is not suitable for ambiguous or exploratory queries where the relevant interpretation of the query semantics among several alternatives is unknown.

There are also works that generate new queries based on popular words in the original query result [24, 7, 5, 22, 15, 21], considering factors like term frequency, inverse document frequency, ranking of the results in which they appear, etc. In particular, [15, 22] exploit relational databases instead of text documents, and [22] only considers term frequency but has the advantage of generating expanded queries without evaluating the original one. [23] additionally considers the proximity to the original query keywords when selecting words from results or corpus to compose new queries. As discussed in Section 1 and shown in Section 5, these approaches emphasize on result summarization, and are not suitable for handling exploratory and ambiguous queries.

[21] addresses a specific application: query expansion when searching products with user ratings. It selects a set of product attributes as expanded terms based on co-occurrence patterns, extreme rating (e.g., product attributes mentioned in highly positive or negative reviews) and consistent rating (e.g., features present in unanimous reviews).

**Faceted Search.** Faceted search provides a classification of the data and enables effective data navigation. There are several approaches for automatically constructing faceted navigation inter-faces given the set of query results, which aim at reducing the user's expected navigational cost in finding the relevant results [8, 14, 16].

Compared with faceted search, our approach is advantageous in two cases: (1) when it is difficult to extract facets, such as searching text documents; and (2) when the query is ambiguous. For ambiguous queries like "apple", "eclipse", etc., different results may have completely different facets. In this case, it is difficult for faceted search to navigate the results and disambiguate the query.

**Cluster Labeling / Summarization.** The goal of cluster labeling is to find a set of descriptive words for each cluster, which summarizes the content of the cluster, and meanwhile differentiates it from other clusters. Some representative works include [6, 19]. A typical way of measuring the desirableness of a term is TFICF, i.e., term frequency and inverse cluster frequency. Finding cluster representatives for structured data has also been studied: [1] assumes each result to be a tuple in a relational database with numerical attributes, and uses the $k$-medoids method to generate a representative for each cluster. As we have discussed in Section 1, while cluster labeling techniques are useful, their goal is different as query expansion, and thus they cannot be directly applied for query expansion. Unlike cluster labeling, the *interaction* of the terms needs to be considered in query expansion, making it a lot more challenging. Furthermore, while cluster labeling quality is typically judged empirically, in this work we propose a quantitative measure of query expansion (i.e. the harmonic mean of the F-measures of the expanded queries).

Compared with existing work, there are several uniqueness of our approach. First, compared with existing query expansion approaches, we generate expanded queries with the aim of presenting a classification of the original query results. This is especially useful for handling exploratory queries and ambiguous queries. Second, our technical contributions focus on how to generate queries with high F-measure given the ground truth of query result. To the best of our knowledge, this is the first study on this problem. Furthermore, unlike existing work that addresses the query expansion problem using heuristics, this work formalizes the problem and quantifies the quality of an approach.

**Result Differentiation.** One of our prior works [18] studied the problem of result differentiation. With the goal of differentiating a set of user selected results, [18] selects a set of feature types defined as (entity, attribute) such that results have different values or value distributions on those feature types. For instance, for two stores that both sell outwear, [18] may choose outwear as a differentiating feature for those stores if one store sells a lot of them while another only sells a few. However, such a choice is not good for the query expansion problem as both stores can be retrieved by keyword "outwear". On the other hand, for the query expansion problem, results of the original query may be significantly different and do not share the same type of features. The technique in [18] that selects feature types shared by all results is generally inapplicable. Furthermore, since a query is generated for each cluster of results, the set of selected keywords/features should *co-occur* in many results in this cluster (but not in many results in other clusters). This challenge is not applicable in [18] for differentiating individual results.

## G. Expanded Queries

The expanded queries generated by each approach for the queries in Table 1 are shown in Figures 8 and 9.

| QW1: San Jose | |
|---|---|
| ISKR | q1: "San Jose, Player, Hockey"<br>q2: "San Jose, Location" |
| PEBC | q1: "San Jose, Player,"<br>q2: "San Jose, Location" |
| CS | q1: "San Jose, sabercat, season, arena "<br>q2: "San Jose, war, california, gold" |
| Google | q1: "San Jose, Attractions"<br>q2: "San Jose, costa rica " |
| Data Clouds | q1: "San Jose, scorer"<br>q2: "San Jose, kyle" |
| F-measure | q1: "San Jose, Player, hockey"<br>q2: "San Jose, location" |

| QW3: CVS | |
|---|---|
| ISKR | q1: "CVS, prince, household"<br>q2: "CVS, code, community"<br>q3: " CVS, southwest" |
| PEBC | q1: "CVS, prince, shop"<br>q2: "CVS, wikipedia"<br>q3: "CVS, southwest" |
| CS | q1: "CVS, station, distribution, retail "<br>q2: "CVS, webster, indiana, settlement"<br>q3: "CVS, system, jike, java" |
| Google | q1: "CVS, careers"<br>q2: "CVS, test"<br>q3: "CVS, caremark" |
| Data Clouds | q1: "CVS, Bull"<br>q2: "CVS, gnuplot"<br>q3: "CVS, Java" |
| F-measure | q1: "CVS, station vma"<br>q2: "CVS, eastern, caremark"<br>q3: "CVS, community" |

| QW5: Eclipse | |
|---|---|
| ISKR | q1: "Eclipse, model, software"<br>q2: "Eclipse, march"<br>q3: "Eclipse, greek" |
| PEBC | q1: "Eclipse, model, environment, automate"<br>q2: "Eclipse, hali"<br>q3: "Eclipse, greek" |
| CS | q1: "Eclipse, core, plugin, offical"<br>q2: "Eclipse, role, origin, video"<br>q3: "Eclipse, greek, ancient, athenian " |
| Google | q1: "Eclipse, mitsubishi"<br>q2: "Eclipse, car"<br>q3: "solar, eclipse" |
| Data Clouds | q1: "Eclipse, core, postfix"<br>q2: "Eclipse, role, task"<br>q3: "Eclipse,pagani" |
| F-measure | q1: "Eclipse, Model, software"<br>q2: "Eclipse, March, related"<br>q3: "Eclipse, greek" |

| QW7: Cell | |
|---|---|
| ISKR | q1: "Cell, express, data"<br>q2: "Cell, biological"<br>q3: "Cell, battery" |
| PEBC | q1: "Cell, express"<br>q2: "Cell, language"<br>q3: "Cell, battery" |
| CS | q1: "Cell, biophosphate, placent, mosaic "<br>q2: "Cell, sumono, yumeka, template"<br>q3: "Cell, battery, kinase, amala" |
| Google | q1: "cell, parts of a cell"<br>q2: "Cell, theory"<br>q3: "Cell, animal" |
| Data Clouds | q1: "Cell, multicellular"<br>q2: "Cell, bit"<br>q3: "Cell, stomach" |
| F-measure | q1: "Cell, express data"<br>q2: "Cell, biological"<br>q3: "Cell, battery" |

| QW2: Columbia | |
|---|---|
| ISKR | q1: "Columbia, University, research"<br>q2: "Columbia, Album"<br>q3: "Columbia, british" |
| PEBC | q1: "Columbia, University, college"<br>q2: "Columbia, Album"<br>q3: "Columbia, Mountain" |
| CS | q1: "Columbia, guillermo, calvo, argentina"<br>q2: "Columbia, essential, toni, bennett "<br>q3: "Columbia, wakaheena, history, highway" |
| Google | q1: "Columbia, country"<br>q2: "Columbia, house"<br>q3: "Columbia, wikipedia" |
| Data Clouds | q1: "Columbia, strong"<br>q2: "Columbia, yakama"<br>q3: "Columbia, light" |
| F-measure | q1: "Columbia, University, research "<br>q2: "Columbia, Album"<br>q3: "Columbia, British" |

| QW4: Domino | |
|---|---|
| ISKR | q1: "Domino, page, long"<br>q2: "Domino, album, produce"<br>q3: "Domino, queen" |
| PEBC | q1: "Domino, page, science"<br>q2: "Domino, album"<br>q3: "Domino, queen" |
| CS | q1: "Domino, restart, pizza, food"<br>q2: "Domino, vocal, album, die "<br>q3: "Domino, french, language, christian " |
| Google | q1: "Domino, game"<br>q2: "Domino, movie"<br>q3: "Domino, records" |
| Data Clouds | q1: "Domino, album"<br>q2: "Domino, pizza"<br>q3: "Domino, sugar" |
| F-measure | q1: "Domino, page science"<br>q2: "Domino, album brand"<br>q3: "Domino, retreival" |

| QW6: Java | |
|---|---|
| ISKR | q1: "Java, Server"<br>q2: "Java, code"<br>q3: "Java, Island" |
| PEBC | q1: "Java, server, web"<br>q2: "Java, aspectj"<br>q3: "Java, island" |
| CS | q1: "Java, blog, microsoft, tool "<br>q2: "Java, view, howard, system "<br>q3: "Java, western, south, parallel" |
| Google | q1: "Java, tutorials"<br>q2: "Java, games"<br>q3: "Java, test" |
| Data Clouds | q1: "Java, nabble"<br>q2: "Java, bit"<br>q3: "Java, room" |
| F-measure | q1: "Java, Server"<br>q2: "Java, Data"<br>q3: "Java, Molucca" |

| QW8: Rockets | |
|---|---|
| ISKR | q1: "Rockets, NBA"<br>q2: "Rockets, launch"<br>q3: "Rockets, iowa" |
| PEBC | q1: "Rockets, NBA"<br>q2: "Rockets, Interior"<br>q3: "Rockets, Built" |
| CS | q1: "Rockets, vernon, maxwell, orlando"<br>q2: "Rockets, israel, dome, missile"<br>q3: "Rockets, built, rhode, singer " |
| Google | q1: "Model, Rockets"<br>q2: "Space, Rockets"<br>q3: "Bottle, Rockets" |
| Data Clouds | q1: "Rockets, cincinnati"<br>q2: "Rockets, anti"<br>q3: "Rockets, target" |
| F-measure | q1: "Rockets, NBA"<br>q2: "Rockets, Interior"<br>q3: "Rockets, Built" |

**Figure 8: Expanded Queries (1)**

| QW9: Mouse | | QW10: Sportsman, Williams | | QS1: Canon, Products | |
|---|---|---|---|---|---|
| ISKR | q1: "Mouse, technique"<br>q2: "Mouse, scientific"<br>q3: "Mouse, cartoon" | ISKR | q1: "Sportsman, Williams, smith, point"<br>q2: "Sportsman, Williams, launch"<br>q3: "Sportsman, Williams, stuart" | ISKR | q1: "canonproducts: category: Camcorders"<br>q2: "canonproducts: category: printer"<br>q3: "canonproducts: category: camera" |
| PEBC | q1: "Mouse, technique"<br>q2: "Mouse, scientific"<br>q3: "Mouse, cartoon" | PEBC | q1: "Sportsman, Williams, smith"<br>q2: "Sportsman, Williams, fire"<br>q3: "Sportsman, Williams, club" | PEBC | q1: "Canonproducts: category: camcorders"<br>q2: "Canonproducts: category: printer"<br>q3: "Canonproducts: category: camera" |
| CS | q1: "Mouse, technique, wheel, interface"<br>q2: "Mouse, birch, hesperian, fossil "<br>q3: "Mouse, cartoon, television, adventure" | CS | q1: "Sportsman, Williams, piano, season american"<br>q2: "Sportsman Williams, alliance, youth, Iraqi"<br>q3: "Sportsman Williams, barker, salem, high" | CS | q1: "canonproducts:category:camcorders"<br>q2: "camera:image resolution:4752 x 3168"<br>q3: "camera:shutter speed:15 - 13,200 sec." |
| Google | q1: "Mouse, Pictures"<br>q2: "Mouse breaker"<br>q3: "Mouse, Pictures of mice" | Google | q1: "Sportsman, Williams, football"<br>q2: "Sportsman, Williams, baseball"<br>q3: "Sportsman, Williams, news" | Google | q1: "Canon, Cameras"<br>q2: "Sony, products"<br>q3: "canon camera products" |
| Data Clouds | q1: "Mouse, mystery"<br>q2: "Mouse, laugh"<br>q3: "Mouse, bush" | Data Clouds | Set1:"Sportsman, Williams, gamebook"<br>Set2:"Sportsman, Williams, highway"<br>Set3:"Sportsman, Williams, kick" | Data Clouds | q1: "Canon, Camera"<br>q2: "Canon, Printer"<br>q3: "Canon, canon wp-dc26 underw" |
| F-measure | q1: "Mouse, technique"<br>q2: "Mouse, scientific"<br>q3: "Mouse, cartoon" | F-measure | q1: "Sportsman, Williams, NBA, boston"<br>q2: "Sportsman, Williams, launch"<br>q3: "Sportsman, Williams, stuart" | F-measure | q1: "Canonproducts: category: camcorders"<br>q2: "Canonproducts: category: printer"<br>q3: "Canonproducts: category: camera" |

| QS2: Networking, Products | | QS3: Networking, Products, Routers | | QS4: TV | |
|---|---|---|---|---|---|
| ISKR | q1: "Networking products: category: routers"<br>q2: "firewalls: vlans: portshield"<br>q3: "Networking products: category: switches" | ISKR | q1: "Routers: Name: integr"<br>q2: "Routers: Name: rangemax"<br>q3: "Routers: Name: linksys" | ISKR | q1: "TV: brand: Toshiba"<br>q2: "TV: brand: LG" |
| PEBC | q1: "networking products: category: routers"<br>q2: "networking products: category: firewalls"<br>q3: "networking products: category: switches" | PEBC | q1: "Routers: Name: integr"<br>q2: "Routers: Name: rangemax"<br>q3: "Routers: Name: linksys" | PEBC | q1: "TV: brand: Toshiba"<br>q2: "TV: brand: Samsung" |
| CS | q1: "Networking products: category: routers"<br>q2: "firewalls:name:d-link dir-130 vpn firewall"<br>q3: "Switches:name:d-link*" | CS | q1: "Routers: RJ-45Ports: 4"<br>q2: "Routers: Features: MAC Filtering"<br>q3: "Routers: Name: linksys" | CS | q1: "TV:display area:26""<br>q2: "TV:name:lg 42lg70" |
| Google | q1: "Social Networking products"<br>q2: "Computer Networking products"<br>q3: "Networking products price" | Google | q1: "Networking, wireless, routers"<br>q2: "Network, routers"<br>q3: "Wood routers" | Google | q1: "TV, guide, products"<br>q2: "TV, electronics"<br>q3: "TV, hair products" |
| Data Clouds | q1: "Networking Products, Switches"<br>q2: "Networking products, Ethernet"<br>q3: "Networking products, firewalls" | Data Clouds | q1: "Cisco 1841 routers"<br>q2: "Cisco 1801 integr"<br>q3: "Cisco 1801 integr" | Data Clouds | q1: "TV, Toshiba"<br>q2: "TV, LG"<br>q3: "TV, TV/DVD combo" |
| F-measure | q1: "Networking products: category: routers"<br>q2: "firewalls: form factor: desktop"<br>q3: "Networking products: category: switches" | F-measure | q1: "Routers: Name: integr"<br>q2: "Routers: Name: rangemax"<br>q3: "Routers: Name: linksys" | F-measure | q1: "TV: brand: Toshiba"<br>q2: "TV: DisplayType: LCD HDTV" |

| QS5: TV, Plasma | | QS6: HP, Products | | QS7: Memory | |
|---|---|---|---|---|---|
| ISKR | q1: "TV: brand: Panasonic"<br>q2: "TV: brand: Samsung" | ISKR | q1: "HPproducts: category: printer"<br>q2: "HPproducts: category: battery"<br>q3: "HPproducts: category: laptop" | ISKR | q1: "Memory: category: harddrive"<br>q2: "Memory: category: flashmemory"<br>q3: "Memory: category: ddr3" |
| PEBC | q1: "TV: displayarea: 42`"<br>q2: "TV: displayarea: 50`" | PEBC | q1: "HPproducts: category: printer"<br>q2: "HPproducts: category: battery"<br>q3: "HPproducts: category: laptop" | PEBC | q1: "Memory: category: harddrive"<br>q2: "Memory: category: flashmemory"<br>q3: "Memory: category: ddr3" |
| CS | q1: "TV: displayarea: 42`"<br>q2: "TV: brand :LG" | CS | q1: "HPproducts: category: printer"<br>q2: "HPproducts: category: battery"<br>q3: "HPproducts: category: laptop" | CS | q1: "Memory: name: cavalry*"<br>q2: "Memory: category: flashmemory"<br>q3: "Memory: category: ddr3" |
| Google | q1: "TV Plasma vs lcd"<br>q2: "TV LCD"<br>q3: "TV, bestbuy plasma" | Google | q1: "HP Products Corporation"<br>q2: "HP Printers"<br>q3: "HP Laptops" | Google | q1: "Human memory"<br>q2: "Computer memory"<br>q3: "Memory game" |
| Data Clouds | q1: "Panosonic"<br>q2: "Samsung"<br>q3: "42`" | Data Clouds | q1: "Battery"<br>q2: "compatible models"<br>q3: "printer" | Data Clouds | q1: "Flash Memory"<br>q2: "DDR2"<br>q3: "DDR3 " |
| F-measure | q1: "TV: brand: Toshiba"<br>q2: "LCD, TV" | F-measure | q1: "HPproducts: category: printer"<br>q2: "HPproducts: category: battery"<br>q3: "HPproducts: category: laptop" | F-measure | q1: "Memory: category: flashmemory"<br>q2: "Memory: category: harddrive"<br>q3: "Memory: category: ddr3" |

| QS8: Memory 8GB | | QS9: Memory Internal | | QS10: Printer | |
|---|---|---|---|---|---|
| ISKR | q1: "Flash Memory: memory size: 8gb`"<br>q2: "Memory: Category: HardDrive"<br>q3: "Memory: Category: DDR3" | ISKR | q1: "Memory: category: Harddrive"<br>q2: "Memory: category: Flash Memory" | ISKR | q1: "Printer: printmethod: laser"<br>q2: "Printer: printmethod: inkjet, |
| PEBC | q1: "Memory: Category: FlashMemory"<br>q2: "Memory: Category: HardDrive"<br>q3: "Memory: Category: DDR3" | PEBC | q1: "Memory: Category: Harddrive"<br>q2: "Memory: Category: FlashMemory" | PEBC | q1: "Printer: Name: imageclass"<br>q2: "Printer: Name: Pixma*" |
| CS | q1: "Flash Memory: memory size: 8gb`"<br>q2: "Memory: category: harddrive"<br>q3: "Memory: category: ddr3" | CS | q1: "Memory: name: hitachi*"<br>q2: "Memory: category: Flash Memory" | CS | q1: "Printer: Condition: New"<br>q2: "Printer: PrintMethod: Laser" |
| Google | q1: "Memory cards 8gb"<br>q2: "Laptop memory, 8GB"<br>q3: "Flash memory" | Google | q1: "dell internal memory"<br>q2: "d internal dell" | Google | q1: "Canon, Printer"<br>q2: "HP, Printer" |
| Data Clouds | q1: "Harddrive"<br>q2: "Flash Memory"<br>q3: "DDR3" | Data Clouds | q1: "flashmemory"<br>q2: "harddrive" | Data Clouds | q1: "pixma"<br>q2: "imageclass" |
| F-measure | q1: "Memory: category: Flash Memory"<br>q2: "Memory, 8GB, Transcend"<br>q3: "Memory: category: DDR3" | F-measure | q1: "Memory: category: Harddrive"<br>q2: "Memory: category: Flash Memory" | F-measure | q1: "Printer: Name: imageclass"<br>q2: "Printer: Name: Pixma" |

**Figure 9: Expanded Queries (2)**